\documentclass[twocolumn,showpacs,prl,amsmath,amssymb]{revtex4}

\usepackage{graphicx}
\usepackage{bm}
\usepackage[varg]{txfonts}

\begin{document}

\title{Griffiths phase in diluted magnetic semiconductors}

\author{V. M. Galitski}
\author{A. Kaminski}
\author{S. \surname{Das Sarma}}
\affiliation{Condensed Matter Theory Center, Department of Physics,
University of Maryland, College Park, MD
20742-4111}

\begin{abstract}
We study the effects of disorder in the vicinity of the ferromagnetic
transition in a diluted magnetic semiconductor in the strongly
localized regime. We derive an effective polaron Hamiltonian, which
leads to the Griffiths phase above the ferromagnetic transition point.
The Griffiths-McCoy effects yield non-perturbative contributions to
the dynamic susceptibility.  We explicitly derive the long-time
susceptibility, which has a pseudo-scaling form, with the dynamic
critical exponent being expressed through the percolation indices.

\end{abstract}


\maketitle

{\em Introduction.} There has been increasing interest recently in
studying diluted magnetic semiconductors both experimentally and
theoretically~\cite{ref1,ref2}.  This interest is motivated not only
by the possibility of potential technological applications of the
materials but also by a very rich physics, which follows from the
unique combination of magnetic and semiconductor properties
co-existing in diluted magnetic semiconductors.  There is a large
number of different materials currently being studied such as ${\rm
  Ga}_{1-x}{\rm Mn}_x{\rm As}$, ${\rm In}_{1-x}{\rm Mn}_x{\rm As}$,
${\rm Ga}_{1-x}{\rm Mn}_x{\rm P}$, ${\rm Ga}_{1-x}{\rm Mn}_x{\rm N}$,
${\rm Ge}_{1-x}{\rm Mn}_x$, {\em etc.} One of the common important
features of these materials is the invariable presence of strong
disorder, which plays an essential role in both magnetic and transport
properties of the systems with the most prominent effect being the
localization of carriers. Disorder arises here from the random
locations of magnetic impurities in the host lattice of the
semiconductor and also from other impurities and defects in the
system. It is important to emphasize that disorder effects are
completely neglected in the continuum virtual crystal approximation
often used in the theoretical studies of these systems.

It is now widely accepted that ferromagnetism in diluted magnetic
semiconductors is due to an indirect interaction between  magnetic
impurities, which is mediated by holes. The effective
magnetic coupling strongly depends upon the ratio
$n_{\textrm{i}}^{1/3} L_{\rm loc}$ of the localization radius and the 
mean distance between the magnetic ions. 
In the strongly localized regime, {\em i.e.} when
$n_{\textrm{i}}^{1/3} L_{\rm loc} \ll 1$, the system may be viewed as
a combination of two static magnetic subsystems (magnetic impurities
and holes) with the corresponding Hamiltonian having the following
form:
\begin{equation}
\label{H}
\!{\cal H} = \sum_{i\,a} 
J  \left( \left| {\bf r}_a - {\bf r}_i \right| \right) 
 {\bf S}_a \bm{ \sigma}_i +\! \sum_{a\,b} J_{\rm AF} \left( \left| {\bf r}_a - {\bf r}_b \right| \right)
  {\bf S}_a{\bf S}_b,
\end{equation}
where ${\bf S}_a$ are impurity spins and $\bm{\sigma}_i$ are hole
spins, $J(r)$ is the manganese-hole interaction, which is given by
\begin{equation}
\label{Mn-h}
J(r)  = J_0 \exp\left(-{r \over L_{\rm loc}} \right),
\end{equation}
and $J_{\rm AF}(r)$ is the direct antiferromagnetic coupling
originating from the exchange interaction of the deep-lying electron
states of magnetic impurities, which decays exponentially with
distance much more rapidly than the manganese-hole coupling
(\ref{Mn-h}).  The Hamiltonian (\ref{H}) leads to a polaron
percolation picture \cite{KaminskiDasSarma02}, which qualitatively
explains the mechanism of the ferromagnetic transition in the system:
In diluted magnetic semiconductors, the concentration of holes is much
smaller than the one of manganese atoms $n_{\textrm{i}}$. If $T \ll
|J_0|$, the magnetic impurities around each localized hole get
polarized forming a bound magnetic polaron. As we lower the
temperature, the polaron size $R_{\rm p}(T)$ increases and polarons
overlap forming finite clusters.  At some threshold temperature
$T_{\rm c}$, an infinite cluster appears signaling the ferromagnetic
transition.  The direct antiferromagnetic coupling cuts off the
polaron growth at the distance $R_* = L_{\rm loc} \ln| { J_0 /
  J_{\rm AF}(n_{\textrm{i}}^{-1/3}) }| $, when the manganese
impurities prefer the antiferromagnetic alignment to the coupling to a
distant hole.  If this distance exceeds the polaron size at the
transition point $R_* \gg R_{\rm p}(T_{\rm c})$, the antiferromagnetic
interaction does not affect the ferromagnetic transition
physics and can be neglected at $T \sim T_{\rm c}$ (we accept
this assumption in the present paper).
    
It is possible to formalize the polaron picture developed in the paper
[\onlinecite{KaminskiDasSarma02}] by deriving the effective
Hamiltonian for interacting polarons. This can be achieved (see below)
by integrating out the manganese spins. The resulting Hamiltonian
(\ref{dHres}) is the ferromagnetic Heisenberg-type Hamiltonian with
small logarithmic ``non-Heisenberg'' corrections, which can be
neglected. The corresponding interaction constant depends upon the
inter-polaron distance, which is the distance between the localization
centers of the corresponding holes. Since the spatial distribution of
the holes is the random Poisson distribution, the ferromagnetic
polaron coupling becomes a random variable with a well-defined
distribution function. This maps the initial problem onto a strongly
disordered ferromagnetic Heisenberg model with randomly placed sites. This
system naturally leads to the well-known Griffiths-McCoy effects
\cite{Griffiths,McCoy} in the disordered phase near the phase
transition. The Griffiths-McCoy singularities are central of the
present Letter.

The original paper of Griffiths \cite{Griffiths} addressed the nature
of the transition in a random Ising ferromagnet, although the main
conclusions are relevant to much more general systems.  The central
idea of the Griffiths theory is that above the ferromagnetic
transition point in a disordered system, there is always a finite
probability of finding an arbitrary large ferromagnetic cluster. These
rare-fluctuation clusters give singular contributions to the
magnetization, which can be proven to be a non-analytic function of
the external magnetic field. The corresponding phase is called the
Griffiths phase.  The statement of the non-analyticity of the
magnetization is a purely mathematical statement. The corresponding
corrections, although singular, are extremely (exponentially) small
and it is not clear whether such effects in thermodynamic quantities
are observable in real experiments.  However, the Griffiths effects
are much more pronounced in dynamic quantities because the disorder in
the model is quenched (does not depend on time) and thus the
rare-fluctuation induced clusters are infinitely extended in the time
direction. Thus, in the Griffiths phase region, the dynamic response
is governed by the Griffiths-McCoy effects as was shown in
Ref.~\cite{Bray}.  Currently there is a great deal of interest in 
Griffiths physics related to quantum phase transitions \cite{QPT,SS}
(a quantum $d$-dimensional system is equivalent to a $d+1$-dimensional
classical systems at $T=0$; thus, a quantum phase transition in a
disordered system is governed by Griffiths-McCoy effects due to
disorder being perfectly correlated in time). 


{\em Effective polaron Hamiltonian.} Let us consider two localized
holes embedded into the system of impurity spins. Let the inter-hole
distance be large: $r \gg n_{\textrm{i}}^{-1/3}$. The partition
function for the system reads:
\begin{eqnarray}
\label{Z}
Z  = {\rm Tr}_{\bm{\sigma}_1, \bm{\sigma}_2} 
 {\rm Tr}_{\left\{ {\bf S_a} \right\}}
\prod_a \exp\Biggl\{- \frac1T\, \hat{\bf S}_a
  \Bigl[&&\!\!\!\!\!   \!\!\!\!\! J\left(\left| {\bf r}_a - {\bf
  r}_1 \right| \right) \hat{\bm{\sigma}}_1 \\
&+& \!\!\!  J\left(\left| {\bf r}_a - {\bf
  r}_2 \right| \right) \hat{\bm{\sigma}}_2  \Bigr] \Biggr\}.
\nonumber 
\end{eqnarray}
It is important to realize that we are building an effective
description in terms of bound magnetic polarons, which are very
``heavy'' classical objects, and the impurity spins are generally
large (5/2 for manganese impurities). Thus, we can ignore quantum effects and
replace the traces in Eq.~(\ref{Z}) by the corresponding classical
integrals. We rewrite Eq.~(\ref{Z}) as
\begin{equation}
\label{Z2}
Z = \oint {d \Omega_1 \over 4 \pi} {d \Omega_2 \over 4 \pi}
\exp\left\{ - \frac{{\cal E}_{\rm eff}\left( \cos\theta, T
\right)}{T}  \right\},
\end{equation}
where integration over $\Omega_i$ implies averaging over the
orientation of the $i$th polaron's spin, $\theta$ is the angle between
the spins of the two polarons, and the effective interaction is defined
as [see Eq.~(\ref{Z})]:
\begin{equation}
\label{Heff}
{\cal E}_{\rm eff}\left( \cos\theta, T \right) = 
- T  \sum_a \ln \left\{ \frac{\sinh[\tilde{J}({\bf
    r}_a)/T]}{\tilde{J}({\bf r}_a)/T} \right\}, 
\end{equation} 
with
\begin{equation}
\label{J12}
\tilde{J}({\bf r}) = J_0 \sqrt{2} \exp\left(-{r_1 + r_2 \over 2 L_{\rm
    loc}}\right) 
\sqrt{ \cosh\left( {r_1 - r_2 \over  L_{\rm loc}} \right) + \cos{\theta}},
\end{equation}
where $r_{1,2}$ are the distances between a magnetic impurity and
polaron centers (localized holes).  Taking into account the fact that the
density of magnetic impurities is high, we can replace the sum in
Eq.~(\ref{Heff}) by the integral:
\begin{equation}
\label{dHeff}
{\partial {\cal E}_{\rm eff}\left( \cos\theta, T \right) \over
  \partial \cos\theta} = 
-  n_{\textrm{i}} \int d^3{\bf r}\, \tanh\left[\frac{
  \tilde{J}({\bf r}_a)}{T} \right] {\partial \tilde{J}({\bf r}) \over
  \partial \cos\theta}.
\end{equation}
In the limit $L_{\textrm{loc}}\ll r$, we have
\begin{eqnarray}
\label{dHres}
{\cal E}_{\rm eff}\left( \cos\theta, T \right)
 = &-&  
  {J_0^2 \over T}
\left( {\frac{\pi}{6}} L_{\rm loc}^2 r n_{\textrm{i}} \right)
  \exp{\left( -{r \over L_{\rm loc}} \right)} \nonumber \\ 
&&\times \ln \left[ f(\theta) {T \over J}
  e^{r \over 2 L_{\rm loc} }\right] \cos{\theta},
\end{eqnarray}
where $f(\theta)\sim 1$ is a function of $\theta$, which is always
non-zero and can be neglected within the logarithmic accuracy. We
conclude that the effective polaron Hamiltonian is the ferromagnetic
exchange Hamiltonian with randomly placed cites and the couplings
between them exponentially decaying with distance. 

A similar random exchange Hamiltonian was considered by Korenblit {\em
  et al.} in Ref.~[\onlinecite{Shklovskii73}], where it was found that
the ferromagnetic transition in the system can be conveniently
described with the help of the percolation theory. Within the
percolation approach, two interacting polarons are considered either
locked in the same direction (if the effective interaction is larger
than temperature: $J_{\rm eff} > T$) or completely uncorrelated in the
opposite limit, $J_{\rm eff} < T$ (which essentially means that the
weak couplings can be set to zero).  This means that we replace the
initial system of randomly interacting polarons by a system in which
the weaker couplings are cut. When decreasing temperature, the number of
strongly coupled polarons increases and at some temperature, an
infinite cluster appears. This temperature is identified with the
ferromagnetic transition temperature of the system.  We would like to
emphasize that the percolation theory is able to establish $T_c$ only
up to a numeric factor of the order of unity.  In the immediate
vicinity of the ferromagnetic transition point, the percolation theory
may fail since it is not able to account for critical fluctuations,
which become increasingly important at $T\to T_c$.  However, there is
a small parameter in our theory, namely
$n_{\textrm{h}}^{1/3}L_{\textrm{loc}}\ll 1$.  The parameter $p$ of the
percolation theory is related to temperature by
\begin{equation}
\label{p}
p(T)\equiv n_{\textrm{h}}^{1/3}L_{\textrm{loc}}\ln\left(\frac{J_0}{T}
\right)\;,
\end{equation}
with $p_c\approx 0.86$. Thus there exists a domain $1\gg
|p-p_c|/p_c\gg n_{\textrm{h}}^{1/3}L_{\textrm{loc}}$, which is outside
the thermodynamic critical region $|T-T_c|/T_c\ll 1$, but still close
to the percolation transition, where we can use the asymptotic results
of the percolation theory.

{\em Griffiths theorem.} Let us prove the existence of the
Griffiths phase in the framework of the percolation theory. In the
percolation picture, we have a number of finite clusters above the
transition point. The concentration of clusters containing $N$
polarons for $N \gg \xi^D$ is \cite{SS,StaufferAharonyBook}
\begin{equation} 
\label{P(N)}
P(N,T)  \propto N^{-\tau} \left( N \over \xi^D \right)^{\tau - \theta}
\exp\left[- A  \left( N \over \xi^D \right)^\zeta \right],
\end{equation}
where $N$ is the number of polarons in a cluster, $\xi \propto \left(
\left| p(T) - p_c \right| / p_c \right)^{-\nu}$ is the dimensionless
correlation length.  Percolation exponents are as follows
\begin{subequations}
\begin{eqnarray}
&&\tau \approx 2.18,\ \nu\approx0.88,\\
&&T  > T_c:\ \zeta=1,\ \theta=3/2,\ D=2,\\
&&T  < T_c:\ \zeta=2/3,\ \theta=-1/9,\ D=3
\end{eqnarray}
\end{subequations}
for the three-dimensional
percolation. 
Constant $A$ in Eq.~(\ref{P(N)}) is a number of the order of unity.

Each polaron has a spin
$S_{\rm p} =S n_{\textrm{i}} v_p$, where $S$ is the spin of a single
magnetic impurity  and $v_p$ is the
polaron volume which is the volume of the sphere of radius $R_{\rm p}
\simeq L_{\rm loc} \ln(J_0/T)$.  The magnetization of the system
in an external magnetic field ${\bf H}$ can be written as follows:
\begin{equation}
\label{M1}
M(H) = S_{\rm p} \sum_N N P(N) \left[ \coth\frac{N S_{\rm p} H}{T} 
- {T \over  N S_{\rm p} H} \right].
\end{equation}
Let us expand this function in Taylor series
\begin{equation}
\label{series}
M(H) = \sum_N a_N H^N\,.
\end{equation}
Following Ref.~\cite{BS}, we obtain
the radius of convergence $R_H$ of the series:
\begin{eqnarray*}
\label{RH}  
R_H &= &\left(  {T \over S_{\rm p}} \right) \lim\limits_{N
    \to + \infty} \left|  {  a_{N-1} \over a_N } \right|\\
& =& \zeta^{1 \over \zeta} \left( {2 \pi T \over S_{\rm p}} \right) \,
\lim\limits_{N \to \infty} N^{-1/\zeta} = 0.
\end{eqnarray*}
Above the percolation threshold, $\zeta=1$ and the series
(\ref{series}) has zero radius of convergence, which implies the
non-analyticity of the magnetization as a function of the external
magnetic field due to rare disorder fluctuations, {\em i.e.} the
essential Griffiths singularity. The corresponding singularity in the
magnetization and susceptibility is however very weak:
\begin{equation}
\label{M(H)}
M(H) \propto \exp\left[ - {\rm const}\, 
\left(\frac{T}{H}\right)  \right],
\end{equation}
which is typical for the Griffiths effects on thermodynamic quantities
in classical systems.

{\em Dynamic susceptibility.} The 
Griffiths physics manifests itself much stronger in dynamic
quantities. The quantity of interest is the dynamic susceptibility,
which is connected with the spin-spin autocorrelation function
\begin{equation}
\label{C(t)}
C(t) = {1 \over V  n_{\textrm{i}}} \sum_a \left\langle {\bf S}_a(t)
{\bf S}_a(0) 
\right\rangle.
\end{equation}
In the Griffiths phase, the disorder-averaged correlator can be
written as follows \cite{Bray,SS}:
\begin{equation}
\label{C(t)2}
C(t) = \sum_N P(N) C_N(t),  
\end{equation}
where 
$C_N(t)$ is a function describing the relaxation of the magnetic
moment of a cluster of size $N$.

The relaxation of the magnetic moment in a cluster is governed by the
Gilbert equation:
\begin{equation}
\label{gilbert}
{\partial {\bf m} \over \partial t}  = \gamma {\bf m} \times
\left\{ - {\partial {\cal F} \over \partial {\bf m}}  - \eta
 {\partial {\bf m} \over \partial t} + {\bf h}(t) \right\},
\end{equation}
where $\gamma$ and $\eta$ are constants corresponding to the
gyromagnetic ratio and Gilbert damping and ${\bf m} = {\bf M} / V_N$,
is the magnetization density with $V_N$ being the volume of the
cluster $V_N = N v_p = N (4 \pi/3) L_{\rm loc}^3 \ln^3 (J_0/ T)$.
Function ${\cal F} ({\bf m})$ is the free energy of the cluster per
volume as a function of the direction of the magnetic moment, and
${\bf h}(t)$ is the fluctuating magnetic field.  Let us note that the
case of $h=0$ and ${\cal F} ({\bf m}) = - {\bf H m}$ yields the
well-known Landau-Lifshitz equations \cite{LLG} with the dimensionless
damping coefficient $\alpha= \eta \gamma m$.  We assume that
anisotropy of the cluster is weak enough so that parameter $\left[ V_N
\left( {\cal F}_{\rm max} - {\cal F}_{\rm min} \right) / T \right]$ is
very small, which corresponds to the superparamagnetic limit.

The random magnetic field ${\bf h}(t)$ is due to thermal fluctuations.
They are supposed to be Gaussian, with the following
white-noise correlation functions:
\begin{equation}
\label{<h>}
\left\langle {\bf h}(t) \right\rangle = {\bf 0}\,\,\, \mbox{and} 
\,\,\, \left\langle h_{\alpha}(t) h_{\beta}(0) \right\rangle
=\mu \delta_{\alpha \beta} \delta(t),
\end{equation}
where Greek indices label Cartesian coordinates. It is possible (see
Ref. [\onlinecite{Brown}]) to connect the constant $\mu$ with the
properties of the cluster and temperature. To do so, one imposes the
natural condition that the stationary solution of the stochastic
Gilbert equation (\ref{gilbert}) subject to the thermal noise
(\ref{<h>}) reduces to the Boltzmann equilibrium
distribution for the magnetization density: ${\bf m} (\theta,\phi)
\propto \exp[ -  V_N {\cal F} (\theta,\phi) / T]$.
This leads to the following expression for the constant in the
correlator (\ref{<h>}): $\mu = 2 \eta T / V_N$.

It is further possible to derive general Fokker-Planck equations for
the distribution function of the magnetic moment $P[{\bf m};t]$ in the
case of an arbitrary anisotropy ${\cal F}({\bf m})$ (this was done in
the pioneering paper of Brown \cite{Brown}).  We however confine
ourself to studying the isotropic case only when the Fokker-Planck
equation reduces to the simple diffusion equation on a sphere (also
described phenomenologically by Bray in Ref.~\cite{Bray}):
\begin{equation}
\label{diff}
\left[ {\partial \over \partial t} - {\cal D} \Delta_{\theta,\phi}
\right] P[{\bf m}, t] = 0,
\end{equation}
where $\Delta_{\theta,\phi}$ is the angular part of the Laplacian and
the effective diffusion coefficient is equal to (see {\em e.g.}
Ref.~[\onlinecite{Brown}])
\begin{equation}
\label{D(N)}
{\cal D}(N,T) = \frac{3}{4\pi}
\frac{1}{S} \left( { \alpha \gamma \over 1 + \alpha^2}
\right)   
{ T  \over \ln^3{J_0 \over T} }
\frac{1}{n_{\textrm{i}} L_{\rm loc}^3 }
{1 \over  N}, 
\end{equation}
where $\alpha$ is the dimensionless damping coefficient introduced in
the previous paragraph, $\gamma$ is the gyromagnetic ratio, and $S$ is 
the magnetic impurity spin.

Let us suppose that the initial magnetization ${\bf m}(t=0)$ of the
cluster is in the $z$-direction, which corresponds to the initial
condition $P[\theta,\phi;t=0] = \delta(\theta)$ for the distribution
function, defining the Green function for the diffusion equation
(\ref{diff}). The corresponding solution reads:
\begin{equation}
\label{green}
P[\theta;t] = \sum\limits_{l=0}^{\infty} \left( l +  {1 \over 2}
\right)
P_l\left( \cos\theta \right) \exp\left[ - {\cal D} l (l+1)  t
\right] \sin\theta,
\end{equation}
where $P_l(\cdot)$ are Legendre polynomials.
The $z$-component of the magnetization  averaged over the
distribution function (\ref{green}) is exactly the function $C_N(t)$
introduced in
Eq.~(\ref{C(t)2}), which describes the relaxation of the magnetic
moment in a cluster of size $N$:
\begin{equation}
\label{CN}
 C_N(t)  = M \exp\left[ - 2 {\cal D}(N,T) t \right],
\end{equation} 

We now proceed to calculate the spin autocorrelation function by
replacing the sum in Eq.~(\ref{C(t)2}) by the integral
\begin{equation}
\label{C(t)3}
C(t) \propto\!\! \int\limits_0^{+\infty}\! dN {N^{1 - \theta}  \over
  \xi^{d\left(\tau - \theta \right)}}  \exp\left[ - A
\left( {N \over \xi^d} \right)^\zeta\!\!
- 2 {\cal D}(N,T) t \right],
\end{equation}
which can be evaluated by the saddle point approximation:
\begin{equation}
\label{C(t)res}
C(t) \propto \left(\frac{p(T)  - p_c}{p_c} \right)^a t^b 
\exp\left[ - g(T)  \left(\frac{p(T)  - p_c}{p_c} \right)^c t^\Delta \right]\;,
\end{equation}   
where, for the sake of brevity, we have introduced the following indices:
\begin{subequations}
\begin{eqnarray}
\label{i1}
\Delta &=& {\zeta \over 1 + \zeta},\\
\label{i2}
c& =& {\nu D \zeta \over 1 + \zeta},\\
\label{i3}
b& =& {2 -\theta - \zeta/2 \over 1 + \zeta },\\
\label{i4}
a& = &\nu D \left[ \tau - \theta - {\zeta \over 1 +\zeta} \left({ 5
  \over 2} -
\theta \right) \right],
\end{eqnarray}
\end{subequations}
and the function in the exponent reads:
\begin{equation}
\label{f(T)}
g(T) = A^{1 \over 1+\zeta} \left[\zeta^{-{\zeta \over 1+ \zeta}} +
\zeta^{{1 \over 1+ \zeta}}\right] \left[2 N {\cal D}(N,T)  \right]^{\zeta
  \over 1+\zeta},
\end{equation}
where the diffusion coefficient is defined by Eq.~(\ref{D(N)})  and
$A$ is the number of the order of unity introduced in Eq.~(\ref{P(N)}).
The explicit values for the parameters above (below) the
transition are: $\Delta = 1/2\,\,(2/5)$, $c \approx 0.88\,\,(1.06)$,
$b \approx 0\,\,(14/15)$, and $a \approx  0.32\,\,(3.29)$.

{\em Conclusion.} We have shown that a diluted magnetic semiconductor
in the strongly localized regime ({\em i.e.}, if $n_\textrm{h}^{1/3} L_{\rm loc}
\ll 1$) must be in the Griffiths phase in the
vicinity the ferromagnetic transition point. The magnetization is a
non-analytic function of the external magnetic field with the
corresponding singularity being exponentially small.  The dynamic
response in the vicinity of the transition point is determined by
rare-fluctuation Griffiths droplets. These droplets correspond to the
percolating clusters in the polaron percolation picture.  The dynamic
susceptibility has the form (\ref{C(t)res}) of the stretched
exponential $\ln{ C(t)} \propto -t^{\Delta}$, with $\Delta$ being
connected to the percolation indices and equal to $1/2$ in the
Griffiths phase and $2/5$ in the
ferromagnetic phase.  
An important consequence of our
work relates to the role of disorder in ferromagnetic semiconductors.
Since disorder is fundamental to the occurrence of Griffiths phase, an
observation of Griffiths physics in diluted magnetic semiconductors
will have significance with respect to the role of quenched disorder
in these systems. In fact, we believe that insulating diluted magnetic
semiconductors ({\em e.g.}, GaMnAs or InMnAs on the insulating side of
the metal-insulator transition) may be an ideal system to look for
signatures of evasive Griffiths physics.

This work was supported by the US-ONR, LPS, and DARPA. V.G.
thanks Olexei Motrunich for very valuable comments and discussions
and for providing a copy of his dissertation. 

\bibliography{griffiths}

\end{document}